# Dirac Metamaterial Assembled by Pyrene Derivative and its Topological Photonics

**Kyoung Hwan Choi[1], Da Young Hwang[1], and Dong Hack Suh[1†]**

*1 Advanced Materials & Chemical Engineering Building 311, 222 Wangsimni-ro, Seongdong-Gu, Seoul, Korea, E-mail: dhsuh@hanyang.ac.kr]=*

**Abstract**

Over the past decade, topology has garnered great attention in a wide area of physics. In particular, it has exerted influence on photonics because carefully engineered photonic crystals and metamaterials can help explore the non-trivial state of materials. In this regard, all dielectric metamaterials with large anisotropy, and dipole and multipole Mie resonators have played an increasingly important role in topological photonics. Advantages of Mie resonators make it possible to quest for non-trivial states in three dimensions and theoretical calculation supports its potential. However, it is very difficult to demonstrate this experimentally because it is hard to make the metacrystal by anisotropic meta-atoms despite much effort. Here we report a Dirac metamaterial for 3D topological photonics. It is implemented by a metacrystal self-assembled by a molecule, HYLION-12 which has both anisotropic polarizability and ring current. As its peculiar properties, it has an exotic optical constant that can be used for the electric and magnetic hyperbolic metamaterial, and the double hyperbolic metamaterial in the ultraviolet region. It also showed 142% of reflectance at 242nm as an amplified reflector and asymmetric transmittance up to 30% through the opaque substrate as a Huygens source under 300nm. Furthermore, it demonstrated various phenomena of topological photonics such as Pancharatnam-Berry and waveguide phase merging, wavefront shaping and waveguide on edges as a 3D topological photonic material. The new strategy using polyaromatic hydrocarbons (PAHs) is expected to be an effective way to realize 3D topological photonics.

**Introduction**

Metamaterials are artificial subwavelength-structured media that exhibit unusual optical properties.[1] They have been developed in terms of materials and functions. As material perspectives, all-dielectric metamaterials have been studied to overcome high dissipation of metal.[2] Active and tunable metamaterials have been explored for various functions.[3] Among them, hyperbolic metamaterials have been succeeded in building exotic electromagnetic space to control light propagation.[4] They exhibit the ultra-anisotropic limit of traditional uniaxial crystal, and one of the principal components of either their permittivity ($\varepsilon$) or permeability ($\mu$) tensor is the opposite sign to the other two principal components.[5-7] The only essential criterion for hyperbolic dispersion is that motion of free electrons is constrained in one or two spatial directions.[4] Therefore, metal-dielectric structures, nanowire arrays with hyperbolic dispersion and associated effects have been demonstrated.[8-10] Although hyperbolic metamaterials have successfully demonstrated, there have been very few reports of implementation for all-dielectric hyperbolic metamaterials.[11-13]

On the other hand, topology – a property of photonic materials that characterizes the quantized global behavior of the wave functions over its entire dispersion band – has emerged, in recent years.[14] It is another indispensable degree of freedom, thus opening a path towards the discovery of fundamentally new states of light and possible revolutionary applications. It has been rooted in the condensed-matter-physics, and various 2D topological photonics[15], such as topological nanophotonics[16], active topological photonics[17] and non-linear topological photonics[18], have been reported. Photonic crystals have played a major role in realizing topological photonic bands, and metamaterials recently have been exploited to achieve topological phases. Realization of electromagnetic duality in bianisotropic metamaterials makes the first topological insulating phase in 2D.[19] And chiral and hyperbolic properties make it possible to demonstrate 3D topological phases.[20] Additionally, metallic helices[21], double-helix[22], gyromagnetic hyperbolic metamaterials[23] and Dirac metamaterials[24] are proposed as a topological photonic platform. Although topological photonics have developed rapidly due to the advances of metamaterials, it is challenging to implement the topological phase in deep-subwavelength conditions.[25]

Here a Dirac metamaterial is demonstrated through the exotic organic crystal. It shows the expected topological properties of double hyperbolic metamaterials.[24] Its organic molecule contains PAHs with different polarizability and ring current in all axial directions. So, this is like a superposition of three different ring resonators. The crystal is built with the PAHs which have coincidence of the normal direction and rotates by $\pi/2$. As a metamaterial, amplified reflection and asymmetric transmission as well as topological photonics such as the merging of Pancharatnam-Berry(PB) and waveguide retardation phases, all direction wavefront modulation and the waveguide on the edge state of crystal are also confirmed in the UV region.

**Preparation of double hyperbolic metamaterial**

To use PAHs as a resonator, HYLION-12 was designed and synthesised to build a three-dimensional structure that forms a metasurface for each layer. According to the basic principle of $\pi$- $\pi$ interaction,[26] molecules substituted with strong electron-donating groups at positions 4,5,9 and 10 of pyrene create skewed dimers. (Fig. S1) A dimer of HYLION-12 can have octahedral coordination sites occurring from four interaction sites with two long alkyl chains in the xy-plane and two interaction sites above and below the $\pi$ orbitals of the dimer, which is a skewed pyrene moiety along the z-axis. The detailed methods for the synthesis and crystallization of HYLION-12 are described in the supplementary information (Fig. S2 to S4). Pyrene, the core of HYLION-12, has anisotropic polarizability.[27] The long and short axis directions of the molecule with polarizability is designated as $\alpha_{ll}$ and $\alpha_{mm}$ and polarizability of

the normal vector direction is $α_{nn}$. (Fig. 1a)

The crystal of HYLION-12 was characterized using powder X-ray diffraction (PXRD). The PXRD patterns of the crystal revealed a high degree of crystallinity. (Fig. 1b) The collected PXRD peaks were indexed to obtain the Miller indices using n-TREOR09.[28] The crystal structure was solved by simultaneous annealing using EXPO2013, and the solutions of the structure were further refined using Rietveld refinement.[29,30] Finally, the structure was modelled using Discovery Studio.[31] (Fig. 1c and d) The PXRD patterns exhibited a primitive orthorhombic structure, designated as $Φ_o$, which was classified under a P222 space group. The cell parameters of $Φ_o$ were determined as a = 37.523 Å, b = 34.476 Å, and c = 9.340 Å (residuals: $R_e$ = 8.532, $R_{wp}$ = 11.747, and $χ^2$ = 1.37). The peak at 19.48° indicated that the (002) face represents the vertical distance between two stacked HYLIONs, which was half the value of parameter c. In the refinement process, the cell parameter c was calculated to be two times larger than the nearest distance. In this case, it was the sum of the intra- and inter-dimer distances can be explained by π–π interactions.[32,33]

To confirm the optical characteristics of $Φ_o$, HYLION-12 is spin-coated onto a Si wafer and soda-lime glass, which was designated as $Φ_{o,spin}$. An X-ray diffractometer was used to characterize the coated phase of HYLION-12. $Φ_{o,spin}$ has an XRD pattern similar to that of $Φ_o$, but no peak at 19.48°, corresponding to the (002) face. (Fig. 1e) To verify the disappearance of it, out-of-plane and in-plane XRD analyses were carried out. The out-of-plane XRD pattern yielded the same results as in the θ/2θ mode (Fig. 1f). Meanwhile, the in-plane XRD pattern only had a peak at 19.48° (Fig. 1g), indicating that the phase was fabricated in the edge-on state. It is not a special case in the spin coating of the PAHs.[34,35]

**Characterisation of optical properties**

An ellipsometer is used to measure the optical properties between 200 and 1200 nm with various angles which are 65º, 70º and 75º. (Fig. 2a) The results were fitted with Gaussian oscillators to enforce the Kramers–Kronig consistency, and ε and μ were obtained from ordinary and extraordinary directions. (Fig. 2b and 2c) As HYLION-12 has bianisotropic polarizability, it seemed suitable to interpret it as a biaxial model. However, in this case, $\Phi_{o,spin}$ is formed through the same number of resonators which have polarizability of $\alpha_{ll}$ and $\alpha_{mm}$, respectively. Therefore, it was interpreted more appropriately as a uniaxial model. The detailed information on the fitting parameters can be found in the supplementary information. (Fig. S5-S7)

Permittivity for ordinary direction ($\varepsilon_{ordinary}$) showed a sharp decline between 200 and 310 nm (Fig. 2b). It drops rapidly to less than 1 at wavelengths shorter than 260 nm, and the lowest value is -0.28 at 220 nm. Permeability for ordinary direction ($\mu_{ordinary}$) also clearly bumped between 210 and 250nm. $\varepsilon_{extra-ordinary}$ and $\mu_{extra-ordinary}$ also showed similar results, but their wavelengths were slightly shifted to a shorter region (Fig. 2c). Because $\Phi_{o,spin}$ has an edge on the state, the optical properties of ordinary and extra-ordinary are very similar.

The optical dispersion of $\Phi_{o,spin}$ can be expressed as the four quadrants of electromagnetic response for ordinary and extra-ordinary. (Fig. 2d and e) The ordinary direction of $\Phi_{o,spin}$ shows the negative-index metamaterial in the vicinity of 210 and 220 nm wavelengths, thereby representing that it can act as a Huygens source.[36,37] Moreover, it can be employed as a prefect reflector near the 230nm wavelength.[38,39] In the extraordinary direction, its characteristic as a Huygens source was observed around 220 nm.

High anisotropic ε and μ are another unique aspect of $\Phi_{o,spin}$ for the ordinary and extra-ordinary directions. (Fig. 2f and 2g) It has high anisotropy of positive ε between 240nm and 260nm wavelengths. More interestingly, it has anisotropic ε with a different sign for the different direction between 210 and 230nm. This means that $\Phi_{o,spin}$ is a hyperbolic metamaterial.[4] In the case of µ, there is no positive anisotropy. But it is also a magnetic hyperbolic metamaterial with the ultra-high anisotropy of a different sign between 210nm and 230nm.[40] This can be caused from the anisotropy, and ring current of PAHs. Also, it implies the expansion of all-dielectric metamaterials using PAHs. Through the ordinary and extraordinary µ, it is confirmed that µ also has hyperbolic characteristics around 220nm.

What is even more surprising is that $\Phi_{o,spin}$ exhibits double hyperbolicity.[24] Once ε and µ of the medium have different signs of diagonal components, their ratio is fixed, the electromagnetic duality is frozen, and the circular polarization light becomes the eigenmode.[25] (Fig. 2h) Furthermore, the ratio of ε to µ in each direction clearly depicts a constant value around between 210 and 230nm wavelengths. (Fig. 2i)

**Amplified reflector**

Reflectance is measured to confirm the optical properties of $\Phi_o$ and $\Phi_{o,spin}$. It acts as an amplified reflector, not a perfect reflector around 230 nm, because it is a perfect reflector and a hyperbolic metamaterial at the same wavelength. Therefore, more than 100% reflectance is expected. It shows reflectance of 147% at 244nm by reflection measurements. (Fig. 3a) Moreover, there were two other peaks with reflectance of over 100%, especially at 212 nm and 282 nm, with reflectance of 131% and 120%, respectively. Also, beyond 300 nm, three weaker reflectance peaks were observed.

When light with a frequency lower or close to the material's bandgap frequency enters a high-index Mie resonator, both the magnetic and electric dipole resonances are excited, causing the resonator to behave like a magnetic dipole (first Mie resonance) and an electric dipole (second Mie resonance).[41] The scattered electric field E at a distance larger than the scatter size, can be approximated as the superposition of electric fields created by several first multipole moments of the scatterer: $E \approx E_p + E_m + E_Q + E_M + \cdots$, where $E_p$, $E_m$, $E_Q$, and $E_M$ are the electric fields created by the electric dipole p, magnetic dipole m, electric quadrupole Q, and magnetic quadrupole M moments, respectively.[42] The principle of resonance of HYLION-12 is not the same as the scattering of a Mie resonator, but the same effect is expected. The resonance of the PAHs creates ring current, so it is very similar to the first and the second Mie resonances. Therefore, reflectance of $\Phi_{o,spin}$ can be fitted with Lorentz oscillators. (Fig. 3b and 3c) Although maximum reflectance occurs at the dipole resonance region of the traditional Mie resonator (Si),[43] $\Phi_{o,spin}$ shows the strongest reflectance at the hexapole.

It is essential to understand that PAHs are molecules whose polarizability varies greatly depending on the shape of the molecule and the substituents.[44] Benzene, the basic unit of PAHs, has no difference in polarizability between vertical and horizontal axes on the plane, but has a difference in the normal direction perpendicular to the plane, thereby creating the dipole moment of benzene. However, when an electron-donating groups such as a hydroxide is substituted at the positions 1 and 4 of benzene, the difference in polarizability occurs in vertical and horizontal directions, where the hexapole is naturally embedded.[27] The stronger hexapole can be made when the resonance ring of PAHs is anisotropically connected, as in anthracene and pyrene.[27] HYLION-12 is based on pyrene which is substituted with a strong electron-donating groups at positions 4, 5, 9 and 10. Therefore, it creates a strong inherent hexapole once unpolarised lights are absorbed and all resonators are vibrated.

**Asymmetric transmittance of the chiral hyperbolic metamaterial to the opaque layer**

According to ellipsometric analysis results, $\Phi_{o,spin}$ shows the characteristics of a perfect reflector, and a Huygens source. This region coincides with that of the hyperbolic metamaterial. The best way to access these characteristics is to analyse asymmetric transmittance (AT). Conventional AT originated from the hyperbolic metamaterials[45], and phase gradient metamaterials[46] that are implemented with metals. Therefore, a decoupling layer must be prepared. However, in the case of $\Phi_{o,spin}$, the cause is different, and it shows the same phenomenon as a Mie resonator, so a decoupling layer for the far field is unnecessary. In addition, it is expected to have characteristics as a Huygens source, and it can be observed as the simplest and most efficient AT phenomenon.

The transmittance of $\Phi_{o,spin}$ on soda-lime glass which is opaque under the wavelength of 300 nm is observed. (Fig. 3d) In the backward direction, where $\Phi_{o,spin}$ is not coated, transmittance is not detected below 300 nm. However, in the forward direction, up to 30% transmittance is observed. The largest reflectance was found at 240 nm, and asymmetric transmission increased as the wavelength was shortened, which is consistent with the ellipsometric analysis. It is clear that $\Phi_{o,spin}$ can be used as a Huygens source between 220nm and 230 nm wavelengths.

Furthermore, circular dichroism(CD) spectroscopic results clearly show the exotic phenomena, namely, the chirality of $\Phi_{o,spin}$ below 300nm(Fig 3e) and the characteristics of the double hyperbolic metamaterial (DHM) addressing the analysis result of the ellipsometer. In short, $\Phi_o$ is considered as a double hyperbolic metamaterial, and properties related to topological photonics in $\Phi_o$ can also be expected to occur. All these characteristics were further confirmed using the Mueller matrix measured by ellipsometry, and the results was that the crystal of HYLION-12 can serve as a linear transmitter, and $\pi/2$ and $\pi/4$ retarders. (Fig. S6)

**Topological properties of double hyperbolic metamaterial**

$\Phi_o$ may be a topological characteristic because it has the constant ratio of $\varepsilon$ to $\mu$ between 210nm and 250nm wavelengths. Nevertheless, it is necessary to first consider where it originates from. Herein, two possibilities for it are suggested: One is that HYLION-12 has two superposition ring resonators vibrating at different frequencies. The other is that this orthogonal resonator accumulates by rotating π/2, while matching the normal vector direction of each molecule. If the linear polarized(LP) electromagnetic waves with frequencies of $v_{ll}$ and $v_{mm}$, having resonance with $\alpha_{ll}$ and $\alpha_{mm}$, are irradiated, all molecules will resonate regardless as their positions. (Figure. 4a) However, as the incidence only with $v_{ll}$ proceeds, molecules at the $M_{2n-1}$ position can resonate, but molecules at the $M_{2n}$ position cannot resonate and vice versa. (Figure. 4b and 4c) Therefore, the incidence going forward should undergo a topological phase transition between the $M_{2n-1}$ and $M_{2n}$. As results, an unidirectional light that is immune to back scattering can occur, and this is expected as a topological phenomenon of double hyperbolic metamaterials.[24] Therefore, $\Phi_o$ is anticipated to exhibit a variety of topological photonics.

The experimental data represent the merging of Pancharatnam-Berry(PB) and waveguide retardation phases. Recently, a metasurface which consists of two pairs of nanofins, has been reported to break the symmetry of photonic spin orbital interactions(SOI) by merging PB and the waveguide retardation phases.[47-49] One of the most important characteristics of PB phase is its geometry.[50] According to the relationship between PB and waveguide retardation phases, the orientation angle θ changes as the crystal rotates. Reflectance (or transmittance) must undergo a phase change due to the geometric rotation, proven through careful experiments with an optical microscope. (Fig. S9) As the crystal of HYLION-12 is rotated, the brightest reflection is rotated along the horizontal direction of the crystal (Fig. 5a-c). This is evidence that PB and waveguide retardation phases were verified by other significant features.

The brightness of reflectance will be related to the orientation of HYLION-12 in the crystal. If the incidence is perpendicular to the $\alpha_{nn}$, all ring currents will occur. If the incidence is parallel, only one resonator will resonate. As shown in Fig. 5a, the $\alpha_{nn}$ is perpendicular to the linearly polarised incidence, and $\alpha_{ll}$ and $\alpha_{mm}$ exhibited resonances with the linearly polarised incidence. However, as the crystal of HYLION-12 rotates, polarizability changed from $\alpha_{ll}$ and

$α_{mm}$ to $α_{nn}$. As a result, the brightest reflectance of $Φ_o$ gradually decreased as the angle of θ increased.

In addition, as the angle of the analyser rotates, all directional wavefront shaping is ascertained.(See supplementary movies 1-3) Because the PB phase is a promising approach for achieving an abrupt phase change utilizing the design of gradient metasurfaces,[51] wavefront shaping due to the PB phase can be confirmed in topological photonics.[52] However, in these cases, directions for wavefront shaping are already determined according to the structure of the metasurface or the metamaterial. All directional wavefront shaping has not been observed so far. Therefore, it shows another possibility as an active topological photonic material of $Φ_o$.

Furthermore, the edge of the crystal is shining brightly regardless of the geometric angle of $Φ_o$ and the direction of the analyser. (Fig. 5d-f) Therefore, this light can be regarded as flowing along the plane of the crystal and in the topological band gap. If the phenomenon is the topological waveguide, this phase transition ensures gapless frequency states at the interface. There must be existing edge states at all frequencies within the gap of the bulk mirrors.[14] The observation incidences through filter cube, U-MWB2(Olympus) and U-MWG2(Olympus), verify it, because they do not have the resonance with HYLION-12. Nevertheless, it demonstrates the waveguide of edge state. The phenomenon means that the PAHs in the ground state have its intrinsic ring current which has been theoretically expected.[53,54]

**Conclusion**

Herein, the double hyperbolic metamaterial can be originated from an aromatic molecule. By determining ε and μ using an ellipsometer, the HYLION-12 crystal could be utilised as an amplified reflector and an asymmetric transmitter when implemented in the existing dielectric metamaterial. These properties are demonstrated with reflectance and transmittance of $Φ_o$. Besides, topological photonics are implemented by the double hyperbolic metamaterial. The merging of PB and waveguide retardation phases is directly ascertained and the waveguide along the edge state of the topological photonic band gap is also demonstrated for the first time in UV and visible regions.

**References**

1   Engheta, N. & Ziolkowski, R. W. *Metamaterials: physics and engineering explorations*.  (John Wiley & Sons, 2006).

2   Jahani, S. & Jacob, Z. All-dielectric metamaterials. *Nature nanotechnology* **11**, 23-36 (2016).

3   Boardman, A. D. *et al.* Active and tunable metamaterials. *Laser & Photonics Reviews* **5**, 287-307 (2011).

4   Poddubny, A., Iorsh, I., Belov, P. & Kivshar, Y. Hyperbolic metamaterials. *Nature photonics* **7**, 948-957 (2013).

5   Smith, D. & Schurig, D. Electromagnetic wave propagation in media with indefinite ε and permeability tensors. *Physical Review Letters* **90**, 077405 (2003).

6   Lindell, I. V., Tretyakov, S., Nikoskinen, K. & Ilvonen, S. BW media—Media with negative parameters, capable of supporting backward waves. *Microwave and Optical Technology Letters* **31**, 129-133 (2001).

7   Belov, P. Backward waves and negative refraction in uniaxial dielectrics with negative dielectric ε along the anisotropy axis. *Microwave and Optical Technology Letters* **37**, 259-263 (2003).

8   Liu, Z., Lee, H., Xiong, Y., Sun, C. & Zhang, X. Far-field optical hyperlens magnifying sub-diffraction-limited objects. *science* **315**, 1686-1686 (2007).

9   Ishii, S., Kildishev, A. V., Narimanov, E., Shalaev, V. M. & Drachev, V. P. Sub-wavelength interference pattern from volume plasmon polaritons in a hyperbolic medium. *Laser & Photonics Reviews* **7**, 265-271 (2013).

10  Noginov, M. *et al.* Controlling spontaneous emission with metamaterials. *Optics letters* **35**, 1863-1865 (2010).

11  Yang, Q. *et al.* Mie-Resonant Membrane Huygens' Metasurfaces. *Advanced Functional Materials* **30**, 1906851 (2020).

12  Liu, M. & Choi, D.-Y. Extreme Huygens' metasurfaces based on quasi-bound states in the continuum. *Nano letters* **18**, 8062-8069 (2018).

13  Lan, C. *et al.* Hyperbolic metamaterial based on anisotropic Mie-type resonance. *Optics express* **21**, 29592-29600 (2013).

14  Lu, L., Joannopoulos, J. D. & Soljačić, M. Topological photonics. *Nature*

photonics **8**, 821-829 (2014).

15   Khanikaev, A. B. & Shvets, G. Two-dimensional topological photonics. *Nature photonics* **11**, 763-773 (2017).

16   Rider, M. S. *et al.* A perspective on topological nanophotonics: current status and future challenges. *Journal of Applied Physics* **125**, 120901 (2019).

17   Ota, Y. *et al.* Active topological photonics. *Nanophotonics* **9**, 547-567 (2020).

18   Smirnova, D., Leykam, D., Chong, Y. & Kivshar, Y. Nonlinear topological photonics. *Applied Physics Reviews* **7**, 021306 (2020).

19   Khanikaev, A. B. *et al.* Photonic topological insulators. *Nature materials* **12**, 233-239 (2013).

20   Gao, W. *et al.* Topological photonic phase in chiral hyperbolic metamaterials. *Physical review letters* **114**, 037402 (2015).

21   Kim, M., Lee, D., Lee, D. & Rho, J. Topologically nontrivial photonic nodal surface in a photonic metamaterial. *Physical Review B* **99**, 235423 (2019).

22   Kim, M. *et al.* Extremely broadband topological surface states in a photonic topological metamaterial. *Advanced Optical Materials* **7**, 1900900 (2019).

23   Chern, R.-L. & Yu, Y.-Z. Chiral surface waves on hyperbolic-gyromagnetic metamaterials. *Optics Express* **25**, 11801-11812 (2017).

24   Guo, Q. *et al.* Three dimensional photonic Dirac points in metamaterials. *Physical Review Letters* **119**, 213901 (2017).

25   Kim, M., Jacob, Z. & Rho, J. Recent advances in 2D, 3D and higher-order topological photonics. *Light: Science & Applications* **9**, 1-30 (2020).

26   Hunter, C. A. & Sanders, J. K. The nature of. pi.-. pi. interactions. *Journal of the American Chemical Society* **112**, 5525-5534 (1990).

27   McKinney, J., Gottschalk, K. & Pedersen, L. The polarizability of planar aromatic systems. An application to polychlorinated biphenyls (PCB's), dioxins and polyaromatic hydrocarbons. *Journal of Molecular Structure: THEOCHEM* **105**, 427-438 (1983).

28   Werner, P.-E., Eriksson, L. & Westdahl, M. TREOR, a semi-exhaustive trial-and-error powder indexing program for all symmetries. *Journal of Applied Crystallography* **18**, 367-370 (1985).

29   Altomare, A. *et al.* EXPO2009: structure solution by powder data in direct and reciprocal space. *Journal of Applied Crystallography* **42**, 1197-1202 (2009).

30  Altomare, A. *et al.* EXPO2013: a kit of tools for phasing crystal structures from powder data. *Journal of Applied Crystallography* **46**, 1231-1235 (2013).

31  Biovia, D. S.   (Release, 2017).

32  Cote, A. P., El-Kaderi, H. M., Furukawa, H., Hunt, J. R. & Yaghi, O. M. Reticular synthesis of microporous and mesoporous 2D covalent organic frameworks. *Journal of the American Chemical Society* **129**, 12914-12915 (2007).

33  Percec, V. *et al.* Transformation from kinetically into thermodynamically controlled self-organization of complex helical columns with 3D periodicity assembled from dendronized perylene bisimides. *Journal of the American Chemical Society* **135**, 4129-4148 (2013).

34  Fritz, S. E., Martin, S. M., Frisbie, C. D., Ward, M. D. & Toney, M. F. Structural characterization of a pentacene monolayer on an amorphous $SiO_2$ substrate with grazing incidence X-ray diffraction. *Journal of the American Chemical Society* **126**, 4084-4085 (2004).

35  Yang, H. *et al.* Conducting AFM and 2D GIXD studies on pentacene thin films. *Journal of the American Chemical Society* **127**, 11542-11543 (2005).

36  Pfeiffer, C. & Grbic, A. Metamaterial Huygens' surfaces: tailoring wave fronts with reflectionless sheets. *Physical review letters* **110**, 197401 (2013).

37  Decker, M. *et al.* High-efficiency dielectric Huygens' surfaces. *Advanced Optical Materials* **3**, 813-820 (2015).

38  Moitra, P. *et al.* Large-scale all-dielectric metamaterial perfect reflectors. *Acs Photonics* **2**, 692-698 (2015).

39  Moitra, P., Slovick, B. A., Gang Yu, Z., Krishnamurthy, S. & Valentine, J. Experimental demonstration of a broadband all-dielectric metamaterial perfect reflector. *Applied Physics Letters* **104**, 171102 (2014).

40  Kruk, S. S. *et al.* Magnetic hyperbolic optical metamaterials. *Nature Communications* **7**, 11329, doi:10.1038/ncomms11329 (2016).

41  Bohren, C. F. & Huffman, D. R. *Absorption and scattering of light by small particles*.   (John Wiley & Sons, 2008).

42  Burrows, C. P. & Barnes, W. L. Large spectral extinction due to overlap of dipolar and quadrupolar plasmonic modes of metallic nanoparticles in arrays. *Optics express* **18**, 3187-3198 (2010).

43  Lewi, T., Iyer, P. P., Butakov, N. A., Mikhailovsky, A. A. & Schuller, J. A. Widely tunable infrared antennas using free carrier refraction. *Nano letters* **15**, 8188-8193 (2015).

44  Hammond, J. R., Kowalski, K. & dejong, W. A. Dynamic polarizabilities of polyaromatic hydrocarbons using coupled-cluster linear response theory. *The Journal of chemical physics* **127**, 144105 (2007).

45  Xu, T. & Lezec, H. J. Visible-frequency asymmetric transmission devices incorporating a hyperbolic metamaterial. *Nature communications* **5**, 1-7 (2014).

46  Ling, Y. *et al.* Asymmetric optical transmission based on unidirectional excitation of surface plasmon polaritons in gradient metasurface. *Optics Express* **25**, 13648-13658 (2017).

47  Pu, M. *et al.* Spatially and spectrally engineered spin-orbit interaction for achromatic virtual shaping. *Scientific Reports* **5**, 1-6 (2015).

48  Zhang, F., Pu, M., Luo, J., Yu, H. & Luo, X. Symmetry breaking of photonic spin-orbit interactions in metasurfaces. *Opto-Electronic Engineering* **44**, 319-325 (2017).

49  Mueller, J. B., Rubin, N. A., Devlin, R. C., Groever, B. & Capasso, F. Metasurface polarization optics: independent phase control of arbitrary orthogonal states of polarization. *Physical Review Letters* **118**, 113901 (2017).

50  Xiao, D., Chang, M.-C. & Niu, Q. Berry phase effects on electronic properties. *Reviews of modern physics* **82**, 1959 (2010).

51  Berry, M. V. The adiabatic phase and Pancharatnam's phase for polarized light. *Journal of Modern Optics* **34**, 1401-1407 (1987).

52  Veksler, D. *et al.* Multiple Wavefront Shaping by Metasurface Based on Mixed Random Antenna Groups. *ACS Photonics* **2**, 661-667, doi:10.1021/acsphotonics.5b00113 (2015).

53  Solomon, G. C., Herrmann, C., Hansen, T., Mujica, V. & Ratner, M. A. Exploring local currents in molecular junctions. *Nature Chemistry* **2**, 223-228 (2010).

54  Hirsch, J. Spin-split states in aromatic molecules and superconductors.

Figures

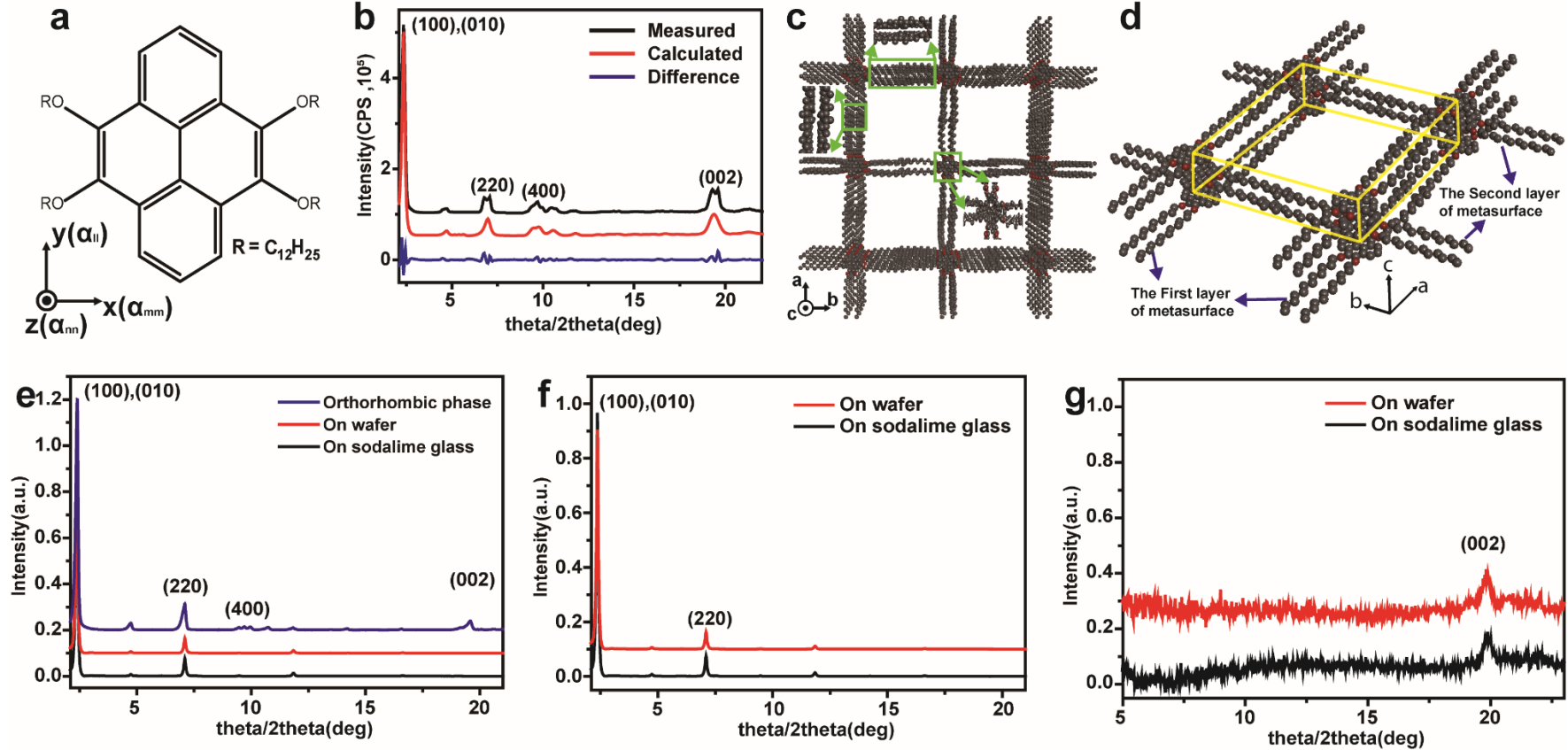

**Figure 1. Structure analysis of HYLION-12. a.** Molecular structure of HYLION-12. Different polarizabilities for three different axes, $\alpha_{mm}$, $\alpha_{ll}$ and $\alpha_{nn}$ for x, y and z axes, respectively. **b.** PXRD patterns of crystal of HYLION-12 and Reitveld analysis; measured (black), calculated (red) difference (blue). **c and d.** Structure modeling and the unit cell of $\Phi_o$, consisted with perpendicular stacked meta-surface. **e.** XRD patterns of $\Phi_{o,spin}$ on wafer (red) and sodalime glass (black). **f and g.** XRD patterns of s $\Phi_{o,spin}$ from out of plane(f) and in plane(g) measurements, respectively.

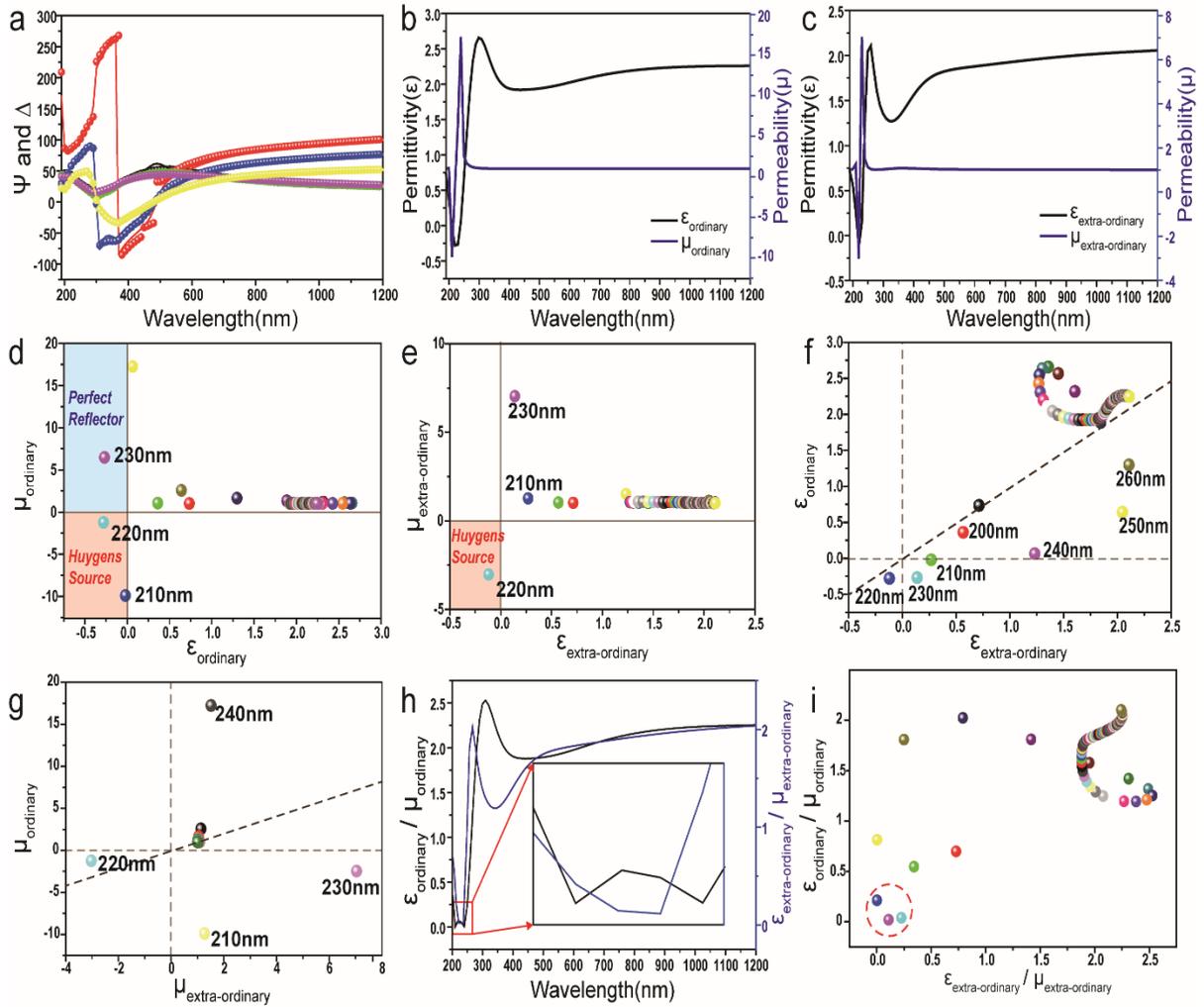

**Figure 2. Optical characteristics of $\Phi_{o,spin}$. a.** Variable Angle Spectroscopic Ellipsometry(VASE) measurements of $\Phi_{o,spin}$ on silicon substrate. Ψ and Δ data are plotted with sphere (experiment) and line (calculation) at different angles of incidence which are 65° (black for Ψ and red for Δ) 70° (green for Ψ and blue for Δ) and 75° (magenta for Ψ and yellow for Δ). **b and c.** ε (black) and μ (blue) of $\Phi_{o,spin}$ for ordinary and extra-ordinary directions. **d and e.** Electromagnetic four quadrant plot for ordinary and extra-ordinary directions. Regardless optical axis, it can be used for perfect reflector or Huygens source depending on wavelength. **f and g.** Anisotropic ε and μ of $\Phi_{o,spin}$. High anisotropy and hyperbolic relation is observed both for ε and μ. **h.** The ratio of ε to μ versus wavelength for ordinary(black) and extra-ordinary(blue). It depicts the constant ratio between 200nm and 250nm wavelengths. This demonstrates $\Phi_{o,spin}$ as double hyperbolic metamaterial. **i.** The anisotropic ratio of ε to μ for extra-ordinary versus ordinary directions.

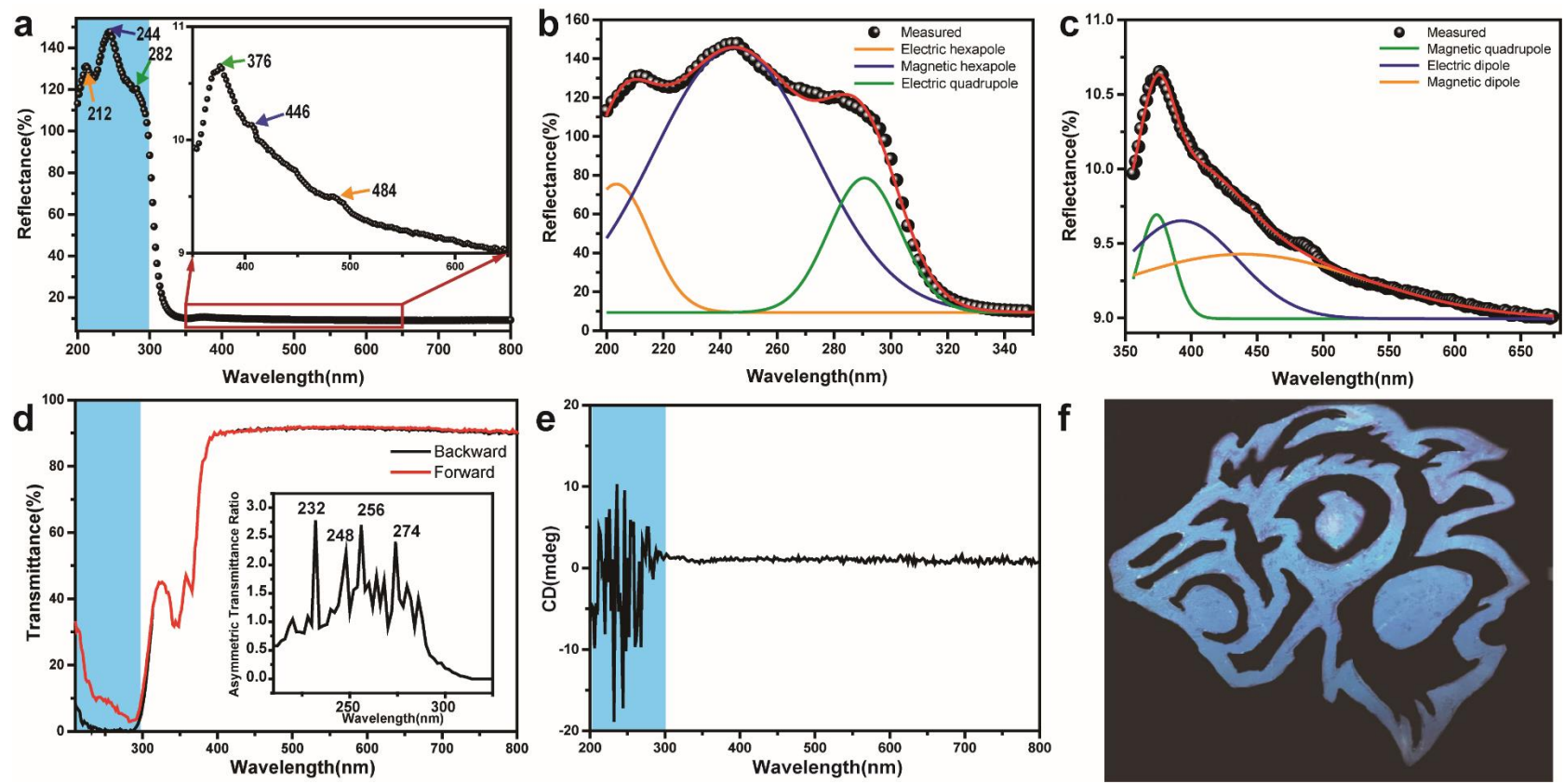

**Figure 3. Reflectance and Asymmetric Transmittance(AT) of $\Phi_{o,spin}$. a.** Reflectance spectrum of $\Phi_{o,spin}$. **b and c.** Contributions of the different multipoles. **d.** Transmission spectra of $\Phi_{o,spin}$ for forward(red) and backward(black) directions. **e.** Measured Circular Dichroism(CD) spectrum of $\Phi_{o,spin}$. Chiral response of $\Phi_{o,spin}$ proved by CD spectrum. **f.** Asymmetric transmittance of $\Phi_{o,spin}$ patterned in CHCl$_3$.

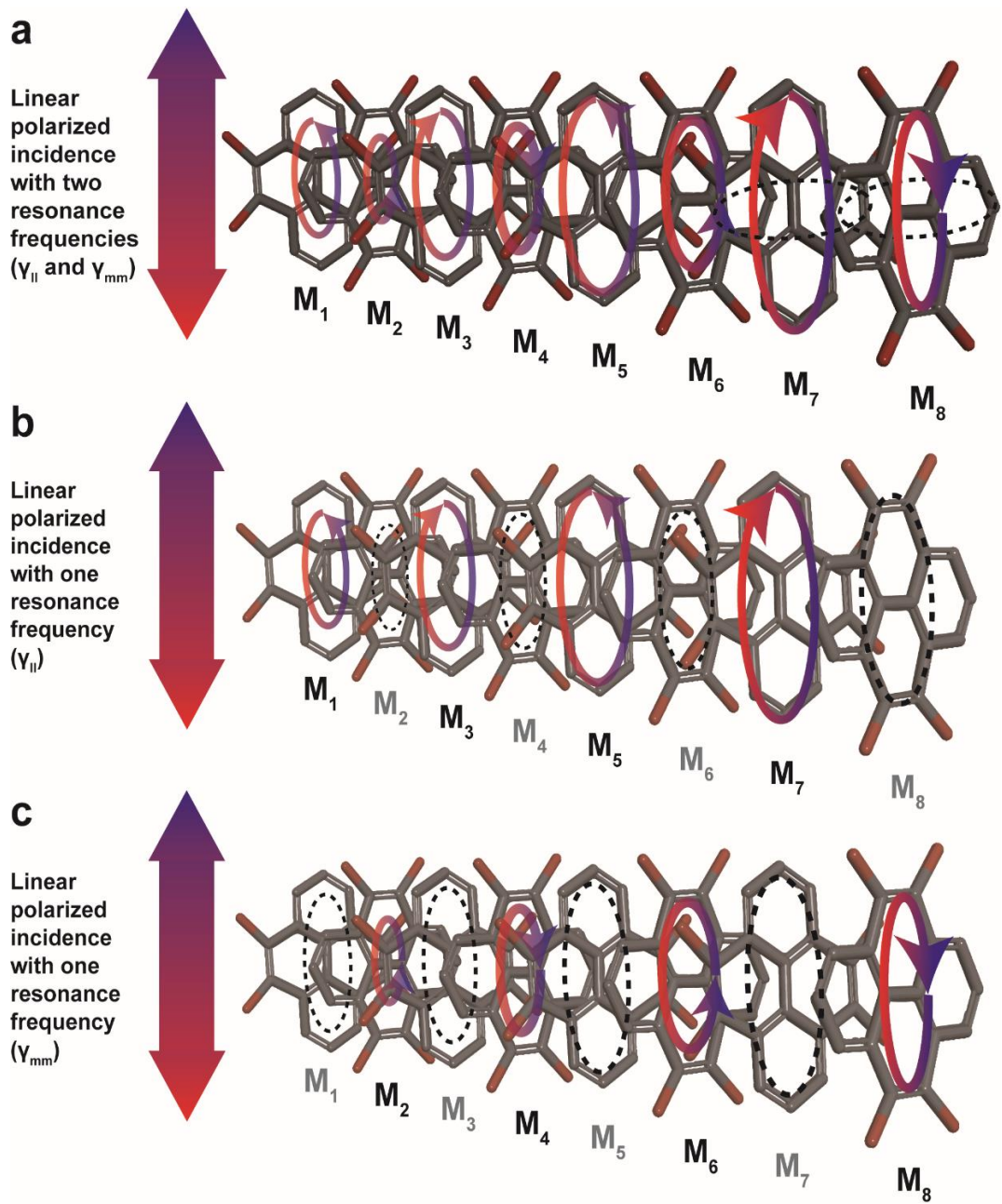

**Figure 4. Scheme of Topological photonics of $\Phi_o$.** HYLION-12 has three superimposed ring resonators, oscillating with different frequencies. **a.** When linearly polarized incidence with two frequencies ($v_{ll}$ and $v_{mm}$) comes, all molecules are resonated with $\alpha_{ll}$ and $\alpha_{mm}$. However, depending on the orientation of HYLION-12, only one of the two resonators in HYLION-12 has a resonating state. For example, molecule at $M_7$ only has resonance with $v_{ll}$ because $\alpha_{mm}$ is rotated with angle of 90°. **b and c.** $M_{2n-1}$ ($M_{2n}$) only resonate with $v_{ll}$ ($v_{mm}$). Therefore, interfacial state and gapless unidirectional waveguide occur.

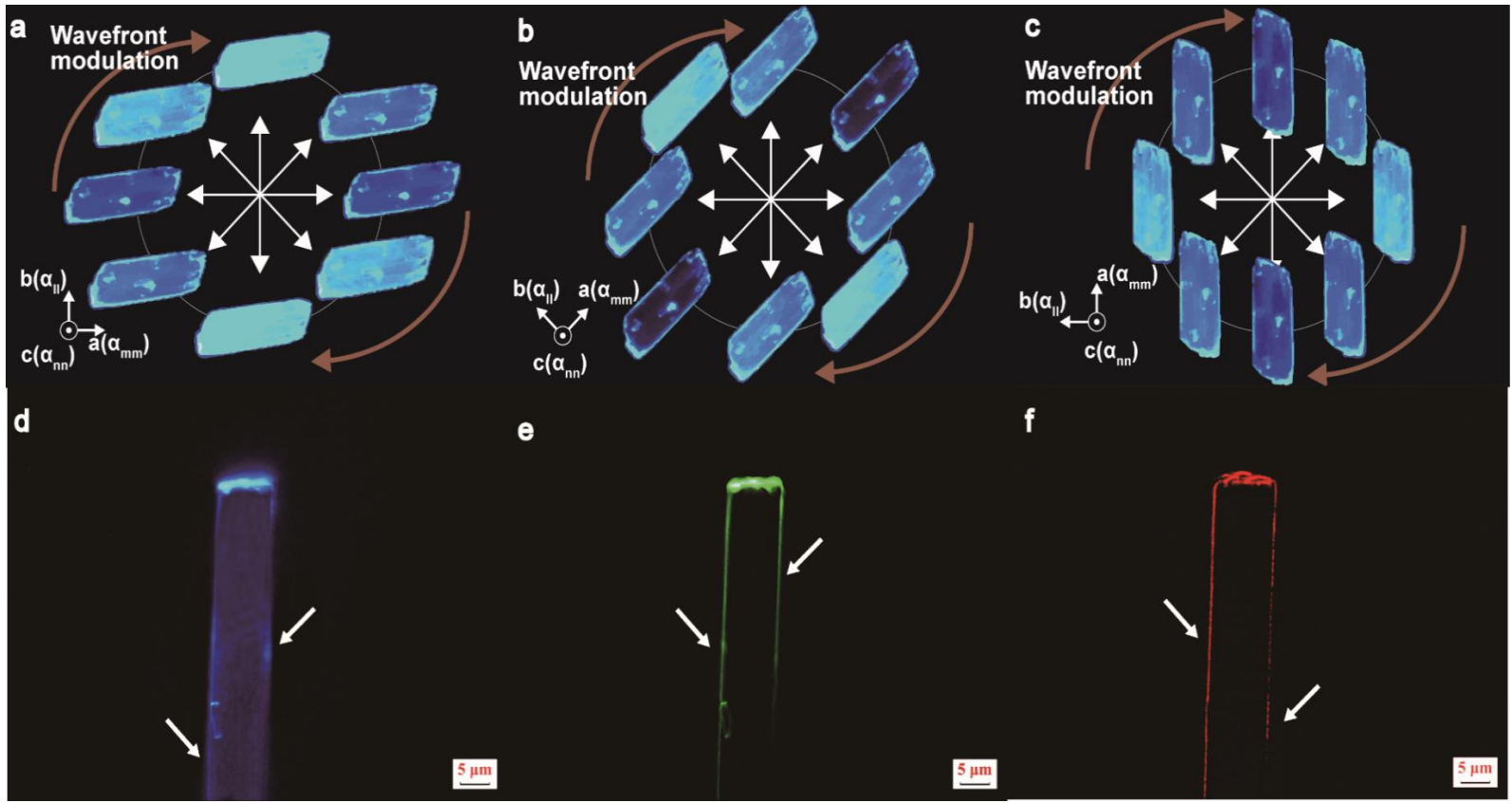

**Figure 5. Topological property of $\Phi_o$. a-c.** Observation of transmittance of $\Phi_o$ with different geometric angles $\theta = 0$, $\pi/4$ and $\pi/2$. It shows the merging between Pancratnam-Berry(PB) and waveguide retardation phases. The brightest states are observed with different directions of analyzer when geometric angles are rotated. **d-f.** Waveguide on edge states of $\Phi_o$. Waveguide on edge states is one of the fundamental topological photonic properties. It occurred at different incidences with filter cube U-MWU2(g), U-MWB2(e) and U-MWG2(f)

## Methods

### Materials

Pyrene, NaIO$_4$, Bu$_4$NBr, Na$_2$S$_2$O$_4$, and Dodecyl bromide were purchased from Sigma Aldrich and used as received. CDCl$_3$ was purchased from Cambridge Isotope Laboratories and was used for the $^1$H NMR spectroscopic studies.

### Synthesis

A mixture of pyrene-4, 5, 9, 10-tetraone(10mmol), Bu$_4$NBr (13mmol), and Na$_2$S$_2$O$_4$ (115mmol) in H$_2$O (50ml) and THF (50ml) was shaken for 5min. Then the color of mixture was changed from dark brown to pale yellow. Bromododecane (60mmol) was added, and followed by aqueous KOH (306mmol, in 50ml H$_2$O). The mixture was stirred for overnight, poured into a mixture solution of H$_2$O (50ml) and ethyl acetate (30ml). The yellow solid was filtered and washed with ethanol. After drying in a vacuum, the solid was recrystallized from ethyl acetate resulting in a white solid with a yield of 85%. (See external data figure 2)

### Characterization of HYLION-12

$^1$H NMR (600 MHz) spectra were recorded on a Varian DRX VNMRS 600 instrument. The purity of the products was determined by a combination of thin-layer chromatography (TLC) on silica gel coated aluminum plates (with F254 indicator; layer thickness, 200 μm; particle size, 2-25 μm; pore size 60Å, SIGMA-Aldrich) Infrared spectra were taken in a KBr disc on a Jasco FT-IR 460 plus spectrometer.

### Crystallization

Crystallization was carried out through gradual cooling and an evaporation method. The HYLION-12 solution was prepared with 1mM with ethyl-acetate and in the case of gradual cooling, the temperature was gradually dropped by 1°C min$^{-1}$ from 50°C to 24°C. In the case of evaporation, the prepared solution was allowed to stand in an 8ml vial at room temperature.

**Spin coating**

To obtain the orthorhombic phase of HYLION crystal coated onto Si-wafer and soda-lime glass, spin-coating was performed with various solvents and concentrations. A mixture of chloroform and ethyl acetate (2/1v) was selected as the best solvent for the spin-coating process. The concentration of HYLION-12, one of the most important variables for spin-coating, was also adjusted between 10 mM to 50 mM, and a solution of 25 mM was identified as the ideal concentration to construct the orthorhombic phase for each substrate.

**X-ray diffractometer analysis**

Theta/2theta, out-of-plane and in-plane x-ray diffraction patterns of all samples were measured on a SMARTLAB (Rigaku Co. Ltd.,) diffractometer using monochromatized Cu-Kα ($\lambda$ = 0.15418 nm) radiation under 40 kV and 100 mA.

**Ellipsometry analysis**

Spectroscopic ellipsometry (SE) was carried out with a Woollam RC-2 ellipsometer from a range of 190 nm to 1800 nm at 65° and 70° light incidence angles and 10 nm steps. Data analysis and fitting were performed with the commercial Woollam CompleteEase software. The measured spectra of the substrate and the crystallized orthorhombic phase was analyzed in a model system composed of two layers. The Si/SiO$_2$ substrate was described using a standard approach with an interface layer between Si and thermal silicon oxide. The metamaterial properties of the orthorhombic phase of HYLION-12 were considered as a mixture of primarily air and HYLION-12, and the characteristic was evaluated using the Anisotropic Bruggeman Effective Medium Approximation (ABEMA). As the optical constants of the metamaterial could be different compared to the bulk materials, parametrization via an analytical function was applied. First, the $\Phi_{o,spin}$ phase was fitted between a range of 1200 nm to 1800 nm with the Cauchy model. After fitting with the Cauchy model in the transparent region, the fitting region was expanded to the short wavelength region with the assumption that it was an absorbing material. The fitting of the psi and del functions was done using an effective medium theory. Finally, the material was modeled as a generic Lorentz and Gauss oscillator between 190 nm to 1800 nm with the uniaxial anisotropic model with an effective medium theory. The final model had three oscillators with an ordinary direction and four oscillators with an extraordinary direction.

**Reflectance and transmittance**

Reflectance and transmittance spectra were measured using a spectrophotometer (UV/vis/NIR V-570, Jasco)

**Dark-field scattering**

The dark-field backward scattering spectra were obtained using a dark-field optical microscope (BX 51, Olympus) integrated with a mercury lamp (U-RFL-T, Olympus), linear polarizer (U-PO3, Olympus), three filter cubes (U-MWU2, U-MWB2 and U-MWG2) analyzer (U-AN360-3) and a CCD camera (NAHWOO).

**Acknowledgement**

This research was supported by Basic Science Research Program through the National Research Foundation of Korea(NRF) funded by the Ministry of Education (2018R1D1A1A02047853). We appreciate the help of Prof. Junsuk Rho(POSTECH) for discussion, Dr. Se Jeong Park (Korea ITS Co. Ltd.) for PXRD and Reitveld analysis were partially executed with her advices and Dr. Ni-Na Hong (J. A. Wollam Co. Ltd.) for ellipsometry data analysis.

**Author contributions**

Dong Hack Suh designed, initiated and directed this research. Kyoung Hwan Choi performed the experiment and analyzed the data. Da young Hwang partially carried out the data analysis and Dong Hack Suh and Kyoung Hwan Choi co-wrote the manuscript. All authors discussed the results.

**Competing interest.** The authors declare no competing interests.

**Supplementary information** is available for this paper

Supplementary information:

# Dirac Metamaterial Assembled by Pyrene Derivative and its Topological Photonics

Kyoung Hwan Choi[1], Da Young Hwang[1], and Dong Hack Suh[1†]

*1 Advanced Materials & Chemical Engineering Building 311, 222 Wangsimni-ro, Seongdong-Gu, Seoul, Korea, E-mail: dhsuh@hanyang.ac.kr*

*Telephone: +82-2-2220-0523, Fax: +82-2-2220-0523*

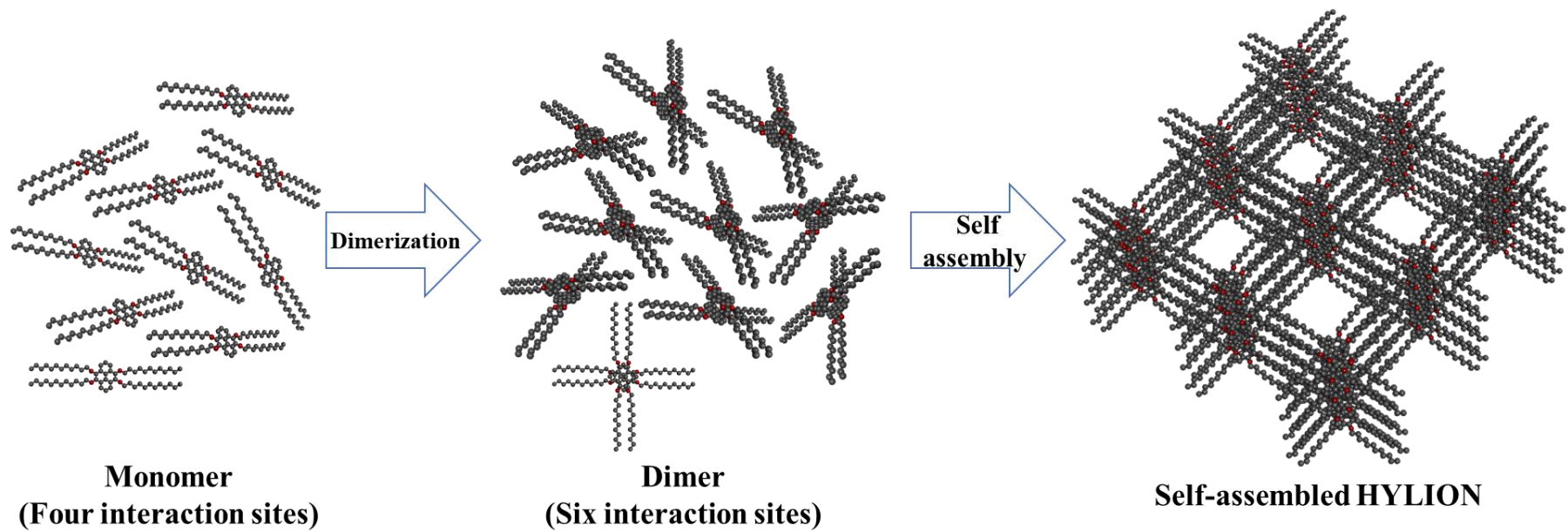

**Figure S1. The Self-assembly process for the periodic and regular structure of HYLION-12.** In general, carbon-based organic molecules have four interaction sites unlike transition metals with d orbital, so it is very difficult or even impossible to build porous three-dimensional structures. To overcome these problems, dimers of HYLION-12 for three-dimensional porous metamaterials were designed to have six interaction sites as a concept of secondary bonds. More specifically, a dimer of HYLION-12 molecule like octahedral molecular geometry has four interaction sites from two long alkyl chains in a plane and two interaction sites from the above and below π orbitals of a dimer which is perpendicularly π-π stacked from pyrene moieties

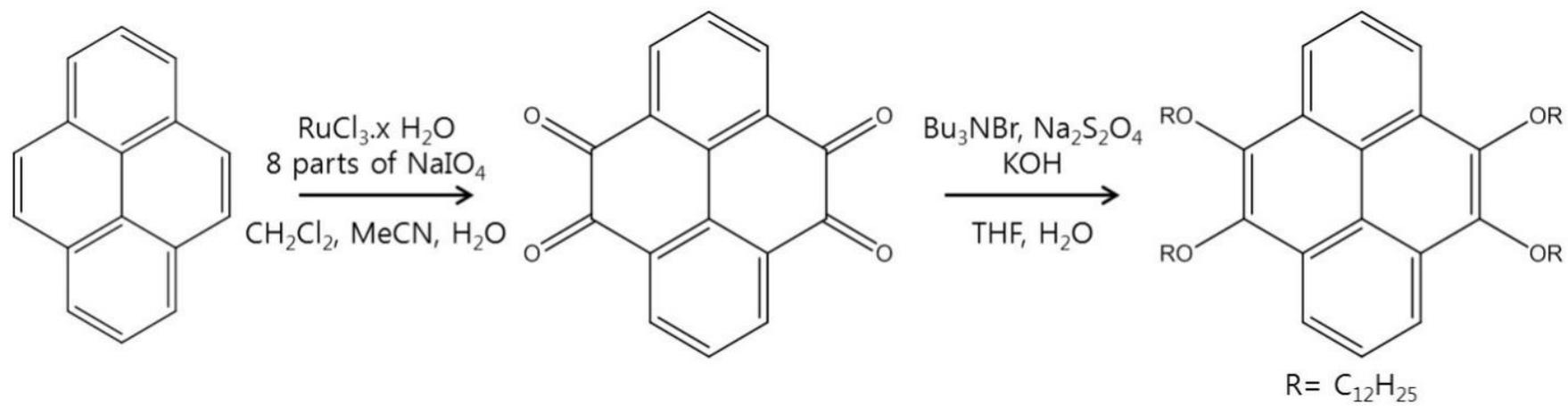

**Figure S2.** Synthesis scheme of HYLION-12(4,5,9,10-tetrakis(dodecyloxy)-pyrenes)

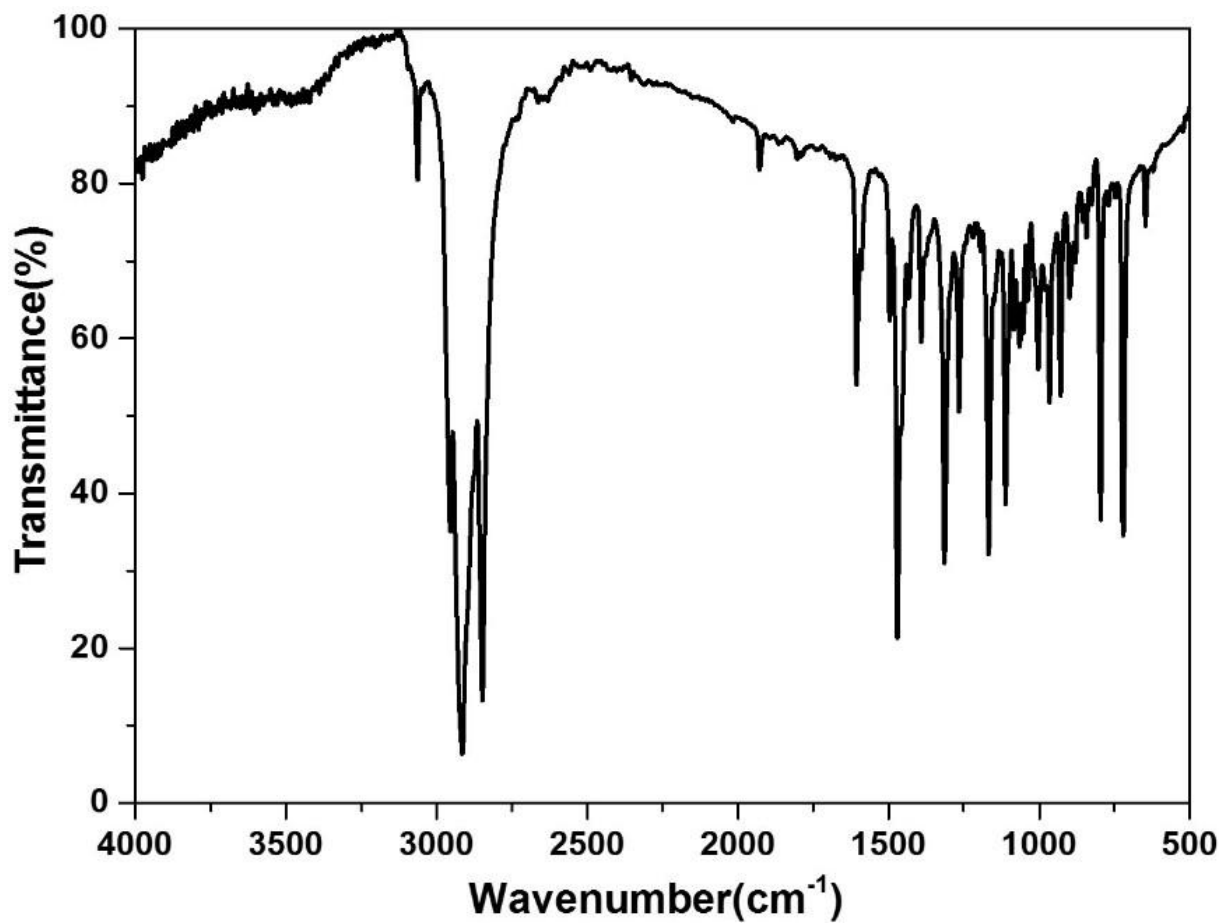

**Figure S3. Infrared spectrrum of HYLION-12.** Infrared spectra of HYLION-12. The characterization of HYLION-12 was fulfilled by Infrared spectroscopy and 1H NMR (600 MHz). The infrared spectroscopy band clearly depict aromatic groups, ether group and alkyl chain stretching for HYLION-12. IR spectra of all HYLIONs well depict aromatic C-H stretching band at 3100cm$^{-1}$, alkyl C-H stretching band at 2918 and 2946 cm$^{-1}$, aromatic C-C stretching band at 1600 and 1493cm$^{-1}$ and ether C-O stretching band at 1267 and 1384cm$^{-1}$

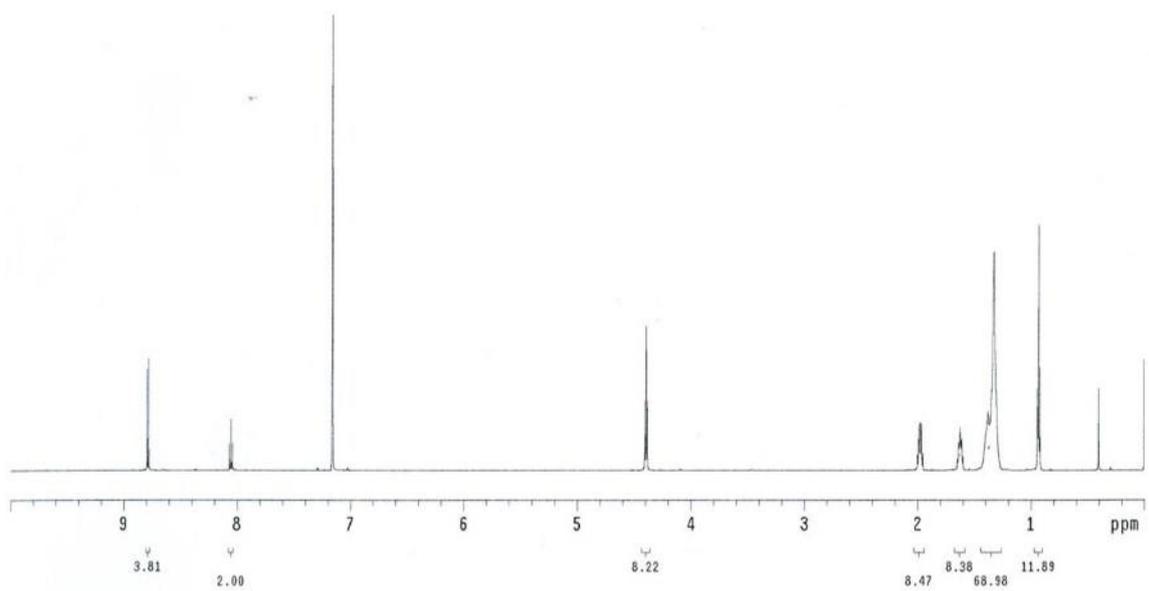

**Figure S4.** The ¹H NMR (600 MHz) specturm of HYLION-12. It is well matched its integration value and peak position for HYLION-12. 1H NMR(600MHz, CDCl3, δ): 8.32(d, 4H), 7.71(t, 2H), 4,21(t, 8H), 1.91(m, 8H), 1.57(m, 8H), 1.40-1.27(m, 64H), 0.88(t, 12H)

```
Layer Commands:  Add  Delete  Save
Include Surface Roughness = OFF
- Layer # 1 = Biaxial   Thickness # 1 = 131.90 nm (fit)
    Type = Uniaxial
    Optical Constants:  Difference Mode = OFF
     - Ex = Gen-Osc
        Add Oscillator   Show Dialog  Fast Gaussian Calc = ON
        Einf = 1.000
        UV Pole Amp. = 0.000  UV Pole En. = 11.000
        IR Pole Amp. = 0.000
        Fit All  Clear All  Add Amp.  Add Br.  Add En.
        1:   Type = Gaussian  Amp1 = 2.668099 (fit)  Br1 = 1.3580 (fit)  En1 = 4.797 (fit)
        2:   Type = Gaussian  Amp2 = 0.278274 (fit)  Br2 = 1.0744 (fit)  En2 = 3.157 (fit)
        3:   Type = Gaussian  Amp3 = 0.589917 (fit)  Br3 = 1.9206 (fit)  En3 = 2.032 (fit)
        4:   Type = Gaussian  Amp4 = 0.995054 (fit)  Br4 = 1.6412 (fit)  En4 = 7.981 (fit)
     - Ez = Gen-Osc
        Add Oscillator   Show Dialog  Fast Gaussian Calc = ON
        Einf = 1.000
        UV Pole Amp. = 0.000  UV Pole En. = 11.000
        IR Pole Amp. = 0.000
        Fit All  Clear All  Add Amp.  Add Br.  Add En.
        1:   Type = Gaussian  Amp1 = 1.914796 (fit)  Br1 = 0.6521 (fit)  En1 = 5.209 (fit)
        2:   Type = Gaussian  Amp2 = 0.352003 (fit)  Br2 = 1.2521 (fit)  En2 = 3.211 (fit)
        3:   Type = Gaussian  Amp3 = 1.115961 (fit)  Br3 = 5.2127 (fit)  En3 = 0.828 (fit)
        4:   Type = Gaussian  Amp4 = 0.465161 (fit)  Br4 = 23.5696 (fit)  En4 = 3.761 (fit)
    Euler Angles:  Phi = 0.00  Theta = 0.00
Substrate = Si_JAW
Angle Offset = 0.00
+ MODEL Options
+ FIT Options
+ OTHER Options
   Configure Options
   Turn Off All Fit Parameters
```

**Figure S5.** Fitting parameter of $\Phi_{o,spin}$.

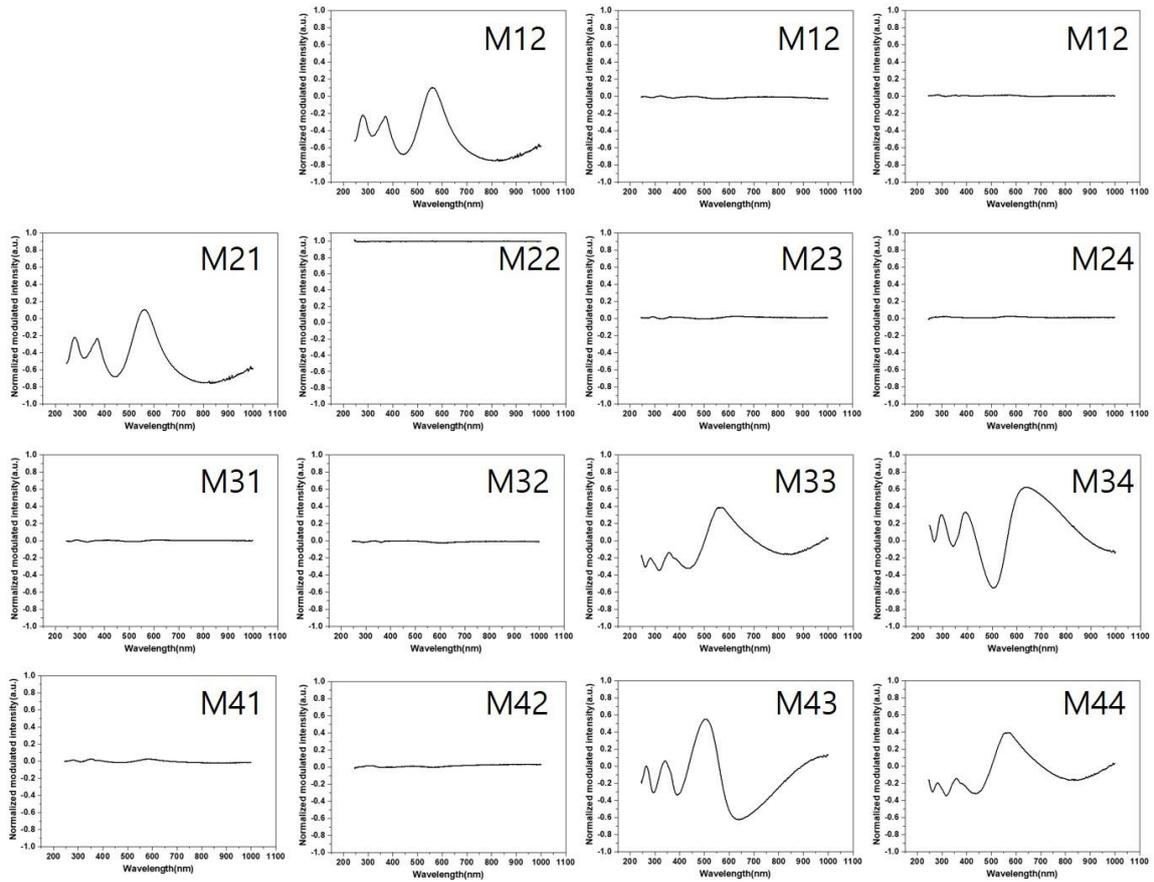

**Figure S6.** Experimental (symbols) and best-model calculated (solid lines) Mueller matrix spectra M12, M13, M21, M22, M23, M31, M32, M33, M41, M42, and M43 obtained for the at angles of incidence is 45º with an in-plane rotation angle is 36º.

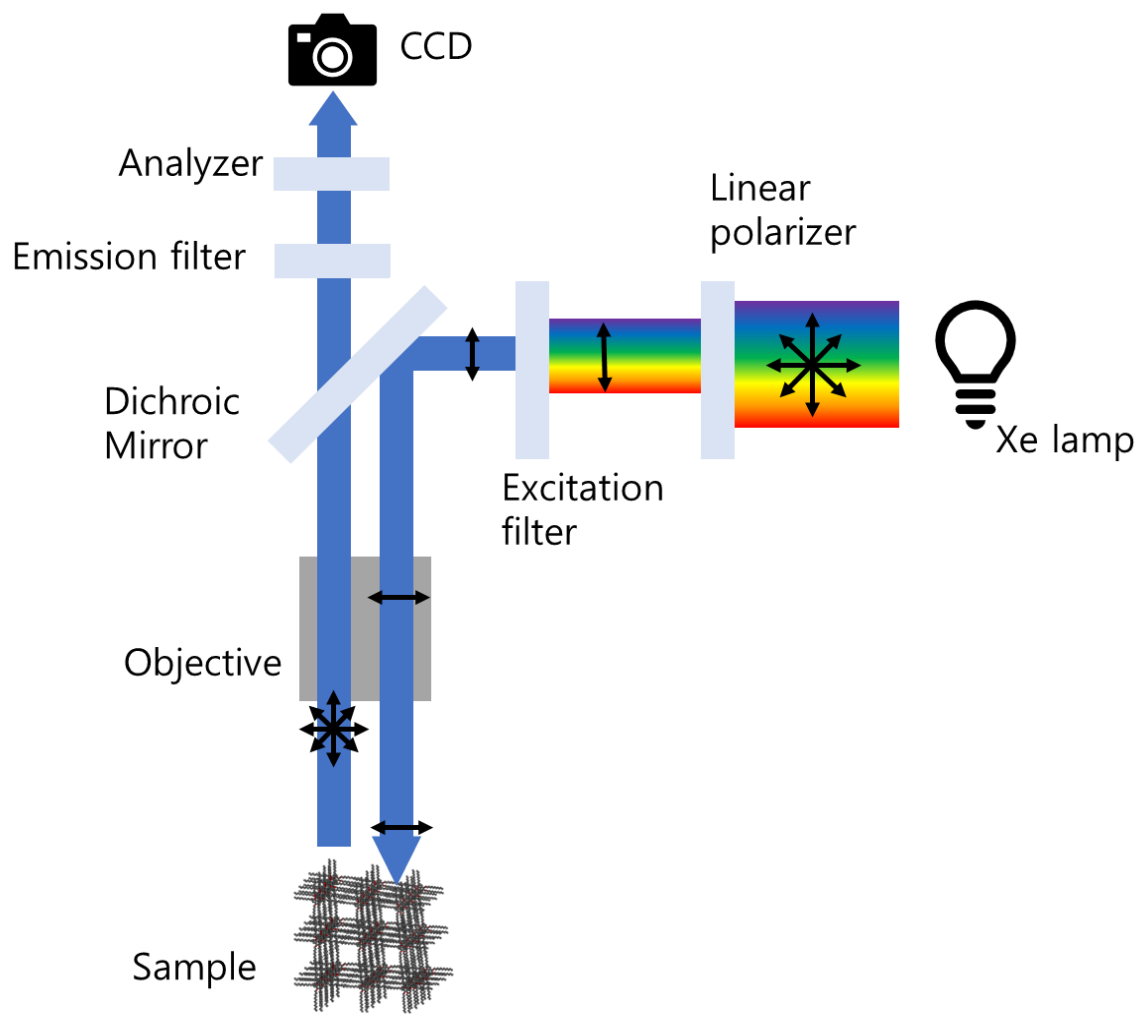

**Figure S7.** Scheme of optical path in dark field scattering.